\documentstyle[prl,aps,epsf,amsfonts,amssymb,multicol,epsfig]{revtex}
 \begin{document}
 \draft
 \title{Nondemolition measurements of a single quantum spin using Josephson oscillations.}
 \author {L. Bulaevskii$^{1}$, M. Hru\v{s}ka$^{1}$, A. Shnirman$^{1,2}$,
 D. Smith$^{1}$, and Yu. Makhlin$^{2,3}$}
\address{$^1$ Los Alamos National Laboratory, Los Alamos, NM 87545, USA \\
$^2$ Instit\"{u}t f\"{u}r Theoretische Festk\"{o}rperphysik
Universit\"{a}t Karlsruhe, D-76128 Karlsruhe, Germany\\
$^3$Landau Institute for Theoretical Physics, Kosygin st. 2,
117940 Moscow, Russia
  }
 \date{\today}
 \maketitle

\begin{abstract}
We consider a Josephson junction containing a single localized
spin 1/2 between conventional singlet superconducting electrodes. We
study the spin dynamics and measurements when a dc-magnetic field
${\bf B }\parallel z$ acts on the spin and the junction is
embedded into a dissipative circuit. We show that when tunneling or a 
voltage are turned on at time $t=0$ the Josephson current starts to oscillate 
with an amplitude depending on the initial ($t=0$) value of the
spin $z$-component, $S_z=\pm 1/2$. At low temperatures,
when effects of quasiparticles may be neglected, this procedure realizes
a quantum-non-demolition (QND) measurement of $S_z$.

\end{abstract}
      
\begin{multicols}{2}

Quantum measurements in mesoscopic systems by use of tunneling 
attracted recently 
great interest. One of the motivations is the challenge 
of the single spin detection~
\cite{BulaevskiiOrtiz,BHO,Zh}. 
The other motivation comes from the quantum computing where 
the final state (after the computation) of a qubit must be 
measured. The meter couples, 
usually, to a single observable of the qubit, e.g., $\sigma_z$.
If this observable commutes with the spin Hamiltonian, the  
QND regime is realized, i.e., the two possible eigenvalues 
of $\sigma_z$ can be measured with proper probabilities even by a
weakly coupled meter. Otherwise the initial state 
is quickly destroyed and one can only observe 
the steady state properties of the qubit and the meter 
performing continuous measurements~
\cite{Averin_Meters,KorotkovAverin,Shnir}.
In the spin detecting tunneling schemes all components of the localized
spin are coupled to the tunneling electrons via the 
exchange interaction. Thus QND measurements seem impossible.

\begin{figure}
\begin{center}
\epsfig{file=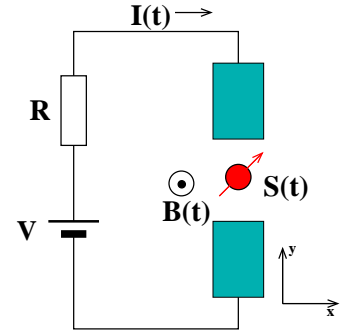,width=0.5\columnwidth,clip=,angle=0}
\end{center}
\caption{The circuit containing Josephson junction with a spin. 
\label{Fig:Circuit} }
\end{figure}

In this paper we show that in fact QND
measurements of the spin projection on the direction of the
applied field ${\bf B }\parallel z$ are possible with the use of
spin dependent Josephson tunneling at low 
temperatures, when the effect
of quasiparticles is negligible. In this case the amplitude of
Josephson oscillations depends on the state of the spin just
before the measurement. The amplitude can be measured, e.g., in a
circuit containing Josephson junction with the spin, a
dc-voltage source $V$ and a resistor $R$, see Fig.~\ref{Fig:Circuit}. The
dissipative spin-dependent current in the circuit appears when the voltage
exceeds a threshold and, effectively, it measures the squared
amplitude of the Josephson-current oscillations.

The measurement can begin either when the voltage exceeding  a threshold 
\cite{note1} or the tunneling in the presence of such a voltage is switched on.
The first possibility appears more natural for mesoscopic
circuits, while the second could be realized with the use of the 
STM with a superconducting tip and a molecule with spin on the
superconducting substrate. As the tip approaches the molecule,
tunneling is turned on.

The system we consider, i.e. a spin in a Josephson junction with
applied voltage, is interesting also from the general viewpoint of
quantum measurements. The entanglement of the spin with the
measurement apparatus is realized here by use of the 
non-dissipative macroscopic quantum system as an intermediate step, 
namely with the superconducting phase degree of freedom. 
Remarkably, in such a system the amplitude of 
Josephson oscillations at frequency $\omega_J=2eV/\hbar$ 
at finite bias $V$ carries information on the initial state
of the spin even for junctions made of singlet
superconductors. Due to the spin conservation in a
singlet-Cooper-pair tunneling in the absence of quasiparticles
($T\rightarrow 0$) the average spin component, $\langle
S_z\rangle$, is preserved to second order in the tunneling amplitude at
$V<2\Delta_0 $ ($\Delta_0 $ being the superconducting gap), though the
spin operator $\hat{S}_z$ does not commute with the Hamiltonian $\hat{{\cal
H}}$ of the system. The noncommutativity of $\hat{{\cal H}}$ and 
$\hat{{\bf S}}$ enables the tunneling measurement of the $z$-component of the 
spin, while due to the preservation of $\langle S_z\rangle$ it is a
QND measurement.

We consider a Josephson junction where the spin-independent tunneling 
is described by amplitude $T_0$ and the spin-dependent tunneling 
by $T_s
\stackrel{<}{\sim} T_0$. The Hamiltonian of the
system is
\begin{eqnarray}
&&{\cal H}={\cal H}_0+{\cal H}_T, \ \ \
{\cal H}_0={\cal H}_a+{\cal H}_b-\mu B_{z}S_z, \\
&&{\cal H}_T=\sum_{n,m,\alpha,\alpha'}a_{n\alpha}^{\dagger}[T_0\delta_
{\alpha,\alpha'}+T_s(\mbox{\boldmath$\sigma$})_{\alpha,\alpha'}\cdot{\bf
S}]b_{m\alpha'}+h.c.,
\label{Eq:H_T}
\end{eqnarray}
where ${\cal H}_a$ and ${\cal H}_b$ are the Hamiltonians of the superconducting
leads a and b, and
$a^{\dagger}_{ n \alpha}$ ($a_{n\alpha }$) creates (annihilates)
an electron in the lead a in the state $n$ with the spin $\alpha$. Further,
$\mbox{\boldmath$\sigma$}$ represents the three Pauli matrices, the 
localized spin 1/2 is described by the operator $\hat{{\bf S}}$, and $\mu$ is the
magnetic moment of the spin. 
In the following we use notations $B$ for $\mu B _z$ 
and $\omega_L  = B /\hbar$.

As we will show in the
following, the Josephson current $I(t)$ oscillates at the frequency
$\omega_J=2eV/\hbar$, and the average oscillation amplitude  depends
on the state of the spin at $t=0$, i.e. on $|\alpha|^2$ for the spin wave
function $\Psi_s(0)=\alpha|\uparrow\rangle+\beta|\downarrow\rangle$.
We show also that the dc-component of the current in the circuit with a 
resistor depends quadratically on the oscillation amplitude, and thus is 
sensitive to the value $S_z$ at time $t=0$.

First, we derive the expression for the Josephson current across
the junction as a functional of the phase difference $\varphi(t)$
across the junction and the spin wave function $\Psi_s(0)$ at low
temperatures, neglecting the effect of quasiparticles, which will be 
estimated later. We neglect quantum fluctuations of the phase
difference across the junction, assuming that the charging energy
$(2e)^2/C$ is lower than the Josephson energy
$\sim T_0^2\rho^2\Delta_0$, where $C$ is the junction
capacitance and $\rho$ is the density of states per spin in the leads. 
We assume that $T_0^2\rho^2,T_s^2\rho^2\ll 1$ and $2eV,B
\ll\Delta_0$, and that all relevant frequencies are much lower
than $\Delta_0 /\hbar$. We find the Heisenberg operator for the
current using the operator perturbation theory \cite{Mahan}
with respect to the tunneling Hamiltonian ${\cal H}_T$, and then we
average this operator over the electron ground state of the
Hamiltonian ${\cal H}_a+{\cal H}_b$. The current operator $\langle
I(t)\rangle_e$ obtained in this way is an operator in the spin space
and $\langle \Psi_s^*(0)|\langle I(t)\rangle_e|\Psi_s(0)\rangle$
defines the average current for any initial state $\Psi_s(0)$. Yet
the measurement produces one of the two values, corresponding to the 
spin up/down states, $S_z=\pm 1/2$, rather than the average.

At time $t>0$ the current is given by
\begin{eqnarray}
&&\hat{I}(t)=\exp(\frac{i}{\hbar}\,\hat{{\cal H}}t)
\hat{ I }\exp(-\frac{i}{\hbar}\,\hat{{\cal H}}t)\ , 
\label{pert1}\\
&&\exp(-\frac{i}{\hbar}\,\hat{{\cal H}}t)=
\exp(-\frac{i}{\hbar}\,\hat{{\cal H}}_0t){\rm T}
\exp\left[-\frac{i}{\hbar}\int_0^t\tilde{{\cal H}}_T
(\tau)d\tau\right], \label{pert2}\\
&&\hat{I}=\sum_{n,m,\alpha,\alpha'}\frac{ie}{\hbar} \hat{a}_{n\alpha}^{\dagger}
[T_0\delta_
{\alpha,\alpha'}+T_s(\hat{\mbox{\boldmath$\sigma$}})_{\alpha,\alpha'}\cdot
\hat{{\bf S}}]\hat{b}_{m\alpha'}+h.c.\label{Eq:Current_Operator}\ ,
\end{eqnarray}
where we accounted for isotropic exchange interaction of tunneling electrons
with localized spin, T is the time ordering operator and
$\tilde{A}(t)=\exp(i{\cal H}_0t/\hbar)\hat{A}\exp(-i{\cal H}_0t/\hbar)$ 
is the operator
in the interaction representation, $\hat{A}$ is the Schroedinger operator and
$\hat{A}(t)$ is the Heisenberg operator.
After averaging over electronic degrees of freedom in the absence 
of quasiparticles and assuming 
a classical phase difference on the junction $\varphi(t)$
we find
\begin{eqnarray}
&&\langle\hat{I}(t)\rangle_e = \frac{4ie}{\hbar^2}
\int_0^tdt'
\sin\left[\frac{\varphi(t)+\varphi(t')}{2}\right]
\times \nonumber \\
&& \left\{F^+(t-t')F(t-t')[T_0^2- T_s^2\tilde{{\bf S}}(t)\cdot
\tilde{{\bf S}}(t')] - h.c.\right\}\ , 
\label{current1}
\end{eqnarray}
where
$\tilde{{\bf S}}(t)=\exp(-i \omega_L \hat{S}_zt){\bf S}
\exp(i \omega_L \hat{S}_zt)$,  
and the Gor'kov Green functions are related to the Bessel functions as
$$
F^+(t)=-F(t)=
(\pi\Delta_0\rho /2)[J_0(\Delta_0 t/\hbar)-iN_0(\Delta_0 t/\hbar)]\ .
$$
For singlet Cooper pairing the dependence of the current
on the spin is isotropic due to the isotropic exchange coupling
of the spin and the tunneling electrons.
We see that the average current  is expressed via the spin correlation
function
$\langle\Psi_s^*(0)|\tilde{{\bf S}}(t)\cdot\tilde{{\bf S}}(t')|\Psi_s(0)
\rangle$.
This function depends on the initial spin state, thus leading to
the corresponding dependence for the supercurrent, because
\begin{equation}\label{SdotS}
\tilde{{\bf S}}(t)\cdot\tilde{{\bf S}}(0)=(2\cos \omega_L t+1)\hat{{\bf 1}}/4-
i \hat{S  }_z \sin \omega_L t,
\end{equation}
where $\hat{{\bf 1}}$ is the unit matrix in the spin space.
The functions $F^+(t)F(t)$, $F^+(t)F(t)\cos \omega_L t$ and 
$F^+(t)F(t)\sin \omega_L t$
oscillate and drop on the scale $\Delta_0^{-1}$ at $B\ll\Delta_0$. For the 
function $\varphi(t)$ with characteristic frequencies well below $\Delta_0$, we
take $\varphi(t)\approx\varphi(t')$ and for $t\gg \Delta_0^{-1}$
we obtain
\begin{eqnarray}
&&\langle\hat{I}(t)\rangle_e={\hat I_c}\sin\varphi(t), \ \ \hat{I}_c=
I_0\hat{{\bf 1}}+I_s\hat{S}_z. \label{current}\\
&&I_0=\frac{2\pi^2e}{\hbar}\left(T_0^2-\frac{3}{4}T_s^2\right)\rho^2\Delta_0,
\ \
I_s=\frac{4e}{\hbar} T_s^2\rho^2B. \nonumber
\end{eqnarray}
$\langle\hat{I}(t)\rangle_e$ is the renormalized current via the Josephson 
junction
obtained after integration over all high frequencies $\gtrsim \Delta_0$. 
It corresponds to the Josephson energy
\begin{equation}
\hat{E}_J\{\varphi(t),\hat{S}_z\}=
(\hbar/2e)(I_0\hat{{\bf 1}}+I_s\hat{S}_z)[1-\cos\varphi(t)],
\label{eh}
\end{equation}
which can be considered as an effective Hamiltonian for the Josephson
junction with the spin.
The spin suppresses the Josephson energy due to the negative contribution
$(3/4)T_s^2$ to
$I_0$ which can lead to the formation of a $\pi$-junction when $I_0$ 
changes sign \cite{bks}. 

The current amplitude depends on the initial state of the spin with 
different values for  
$S_z=1/2$ and $S_z=-1/2$ until the spin 
flips due to quasiparticles or due to relaxation. 
Otherwise, on average, the spin does not flip in Josephson tunneling 
because the first electron of the Cooper pair
may flip it, but then the second one restores the spin since the Cooper pair
cannot carry spin; thus $S_z$ on average is an integral of motion.

The effect of the Josephson current on the spin can also 
be understood from the following argument. 
One sees from Eq.~(\ref{Eq:H_T}) that the spin feels the fluctuating
exchange field induced by tunneling electrons,
\begin{equation}
\label{Eq:Effective_h}
\mu\hat{{\bf h}}(t)=T_s\sum_{n,m,\alpha,\alpha'}a_{n\alpha}^{\dagger}(t)
(\mbox{\boldmath$\sigma$})_{\alpha,\alpha'}b_{m\alpha'}(t) +h.c. \ .
\end{equation}
This field causes deviations of spin from the $z$-axis 
orientation due to presence of transverse components 
$\sigma_x$ and $\sigma_y$ when the first electron of a Cooper pair tunnells 
via spin. 
This spin response interferes with the spin dependent part of the current 
operator (\ref{Eq:Current_Operator}) giving rise to the spin dependent 
contribution in the current. 

Having obtained linear dependence of the Josephson critical current on the 
spin $z$-component we have reduced the problem to the well studied 
case of a dc-SQUID measuring a flux qubit \cite{Averin_Meters}. Here, for 
completeness, we provide a simplified version of the derivation 
of the dephasing and measurement times, valid for $R \ll h/e^2$ and $T \ll V$. 

Without the resistor ($R=0$) the system is dissipationless, 
$\varphi(t)=\omega_J t$ 
and no measurement 
is performed. Introducing dissipation allows us to measure the amplitude of 
the Josephson-current oscillations, and, thus, the initial state of the 
spin. We consider the circuit in Fig.~\ref{Fig:Circuit} with $R \neq 0$.
The circuit dynamics is described by  
\begin{equation}
\frac{\hbar}{2e}C\ddot{\varphi}+\frac{\hbar}{2e}\frac{\dot{\varphi}}{R}+
I_c\sin\varphi=\frac{V}{R}+\xi(t)\ ,
\label{Langevin}
\end{equation}
where the Langevin force 
$\xi(t)$ is the random current (Nyquist noise) 
generated by the resistor. Let us consider the simplest
case when the junction capacitance can be neglected. This is
possible if $RC \omega_J \ll 1$. When $V \ll I_cR$ no voltage drops on 
the junction and the dc-current is not sensitive to the value of the 
Josephson critical current and, thus, to the state of the spin.
For $V\gg I_cR$, solving Eq.~(\ref{Langevin}) and averaging one obtains  
$\varphi(t)=\omega_J t+(I_cR/V)\cos(\omega_J t)$ (see e.g.,~\cite{Koch_Noise}) and 
\begin{equation}
I=I_c\sin\varphi\approx I_c\sin (\omega_J t)+(I_c^2R/V)\cos^2(\omega_J t)\ . 
\end{equation}
The dc-component of the current is given by
\begin{equation}
\hat{I}_{dc}=\frac{\hat{I}_c^2R}{2V}=\frac{(I_0^2+I_s^2/4)R}{2V}\hat{{\bf 1}}+
\frac{I_0I_sR}{V}\hat{S}_z\ .
\end{equation}
Now the dc-current is sensitive to the state of the spin.
For a single measurement one of the two possible values of the current, 
$(I_0\pm I_s/2)^2R/2V$, is
realized. In a multiple set of
measurements these values occur with probabilities
$P_{+}=|\alpha|^2$ and $P_{-}=1-|\alpha|^2$.

Now we are in a position to derive the measurement rate $\Gamma_m$
needed to resolve two values of the current
corresponding to $S_z=\pm 1/2$ on the background of shot noise of 
Cooper pairs tunneling incoherently due to the resistor. 
The signal is $\delta I \equiv  I_0I_sR/V$. 
The noise power of the background current 
$S_I(\omega)$ is defined as
\begin{equation}
S_I(\omega)=\frac{1}{2}\int_{-\infty}^{\infty}dt\exp(i\omega t)\langle
[I(t)I(0)+I(0)I(t)]\rangle\ .
\end{equation}
At $V\gg I_c R$ it is given by
\begin{equation}
S_I(\omega)\approx 2eI_{dc}\approx eI_0^2R/V, \ \ \omega<V\ll\Delta_0,
\label{ii}
\end{equation}
assuming $I_0\gg I_s$. This expression can be obtained~\cite{Koch_Noise} 
from the Langevin equation (\ref{Langevin}) with the appropriate 
high frequency 
spectrum of $\xi(t)$~\cite{AES}. It describes the shot noise of 
individual Cooper pairs tunneling incoherently through the 
junction and dissipating each energy $2eV$ 
into the microscopic modes of the resistor~\cite{Averin_Incoherent_CPT}.
We define the measurement rate as:
\begin{equation}
\Gamma_m \equiv (\delta I)^2/(8
S_I)\approx (I_s^2R)/(8eV) \ .
\end{equation}
Note that the current correlation
function for the shot noise is
expressed via the phase correlation function as
\begin{equation}
\langle I(t)I(0)\rangle=I_0^2\,\langle \sin\varphi(t)\sin\varphi(0)\rangle
\ .
\label{cc}
\end{equation}

Next we discuss the back-action effect of the measurement, i.e. the 
effect of incoherent Cooper pair tunneling, on the spin dynamics.
It leads to the dephasing of transverse spin components, i.e. to 
the destruction 
of coherent superposition of states $S_z=\pm 1/2$. To derive the 
dephasing rate, $\Gamma_d$, we consider the $x$-component of spin
averaged over the electronic degrees of freedom. 
We obtain
\begin{equation}
\label{Eq:Extra_Phase}
\langle S_x(t)\rangle_e = \eta(t) \tilde {S}_x(t)+
\frac{\hbar}{2e}\, I_s\,\tilde{S}_y(t)\int_0^t
\cos\varphi(\tau)d\tau \ ,
\end{equation}
where $\eta(t) < 1$ is an oscillating function close to unity 
[see Eq.~(\ref{eta})], describing 
the reduction of the spin amplitude. 
Its origin will be discussed later. Here
we focus on the second term in the right hand side. This term 
describes the phase accumulation 
due to the effect of a random magnetic field
acting on spin, $h_{zr}(t)=[\hbar I_s/(2e)]\,\cos\varphi(t)$, in agreement with
the effective Hamiltonian (\ref{eh}). 
The dephasing rate is given by
\begin{equation}
\Gamma_d= \frac{1}{2\hbar^2}\,\left(\frac{\hbar I_s}{2e}\right)^2
\int_{-\infty}^{\infty}d\tau
\langle\cos\varphi(\tau)\cos\varphi(0)\rangle\ .
\end{equation}
By use of Eqs.~(\ref{ii}), (\ref{cc}) we get
$\Gamma_d\approx I_s^2R/(8eV)\approx \Gamma_m$.

The inequality $\Gamma_d\geq
\Gamma_m$ is the fundamental property of quantum mechanics as it follows 
from the uncertainty relations (see, e.g., 
review \onlinecite{Devoret_Schoelkopf} and references therein). 
It was proven for the case of a
linear detector with the direct coupling of the detector 
to the measured observable, see e.g., Ref.~\onlinecite{Averin_Meters,Clerk}. 
In our setup the interaction of
spin with tunneling electrons is nonzero only in the second order of
perturbation theory. However, effectively we obtain a linear relation
between the spin $z$-component and the superconducting current 
accessible experimentally, see Eqs.~(\ref{current}), (\ref{eh}).
Consequently, the optimal relation $\Gamma_d\approx \Gamma_m$ is
satisfied for our measurements.

Now we discuss the effects which are beyond the accuracy of the 
effective Hamiltonian (\ref{eh}). 
We calculate the $z$-component of the spin, $\langle S_z(t)\rangle_e$
averaged over the electronic degrees of freedom,
as was done in Eqs.~(\ref{pert1}), (\ref{pert2}). 
We derive for $t\gg \hbar/\Delta_0$:
\begin{eqnarray}
&&\langle \hat{S}_z(t)\rangle_e =
\langle\exp(i{\cal H}t)\hat{S  }_z\exp(-i{\cal H}t)
\rangle_e= \eta(t)\hat{S}_z,\label{cor} \nonumber \\
&&\eta(t)=1-2T_s^2\rho^2[1+\cos\varphi(t)]\ .
\label{eta}
\end{eqnarray}
The maximum average value of the spin $z$-component is slightly smaller
than $1/2$. This is because 
spin flips can happen after one electron of a Cooper pair passes the spin.
The time of passing is $\sim \hbar/\Delta_0$ and probability 
to scatter on the spin per
unit time is $\sim T_s^2\rho^2(\Delta_0/\hbar)\cos\varphi(t)$. 
Also the second term in Eq.~(\ref{Eq:Extra_Phase}), 
which corresponds to an extra phase in the spin rotation, can be understood 
in terms of the spin flips. 
Each passage of a Cooper pair changes this phase
by $\Delta\alpha\sim B/\Delta_0$ and in the case of coherent tunneling
during the time $t$ this phase changes by
$\alpha(t)\sim T_s^2\rho^2(\Delta_0/\hbar)(B/\Delta_0)t\sim T_s^2\rho^2 
\omega_L t$.
Hence, in addition to $B$ there is the effective magnetic field acting on spin $h_z\sim
T_s^2\rho^2B$ as follows also from the effective Hamiltonian (\ref{eh}).
For uncorrelated tunneling of Cooper pairs this field is
random and in addition to regular change it leads to phase diffusion as
$\langle[\alpha(t)-\langle \alpha (t)\rangle]\alpha(0)\rangle\propto
(\Delta\alpha)^2t\propto B^2t$ and $\Gamma_d\propto B^2$.

For comparison, we consider a normal tunnel junction with unpolarized 
electrons in the leads. We find the current sensitivity to the spin $z$ 
component also, 
\begin{eqnarray}
&&\langle \hat{I}(t)\rangle_e  =\frac{2\pi e}{\hbar}\rho^2[(2T_0^2+3T_s^2)V
\hat{{\bf 1}}- \\
&&T_s^2(|V+B|-|V-B|)\hat{S}_z]. \nonumber
\end{eqnarray}
Now, however, the spin flips are allowed because 
tunneling electrons are not paired. 
For short times
$\hbar[T_s^2 \rho _0^2 V ] ^{-1} \gg t \gg  \hbar V^{-1}$
we obtain
\begin{equation}
\langle \hat{S}_z(t)\rangle_e =\hat{S}_z-2\pi T_s^2\rho^2
[2\hat{S}_z{\rm max}(|V|,|B|)-B ]t,
\end{equation}
which gives at $eV\gg B$ the spin-flip (demolition) rate 
$\Gamma^{\dagger}_n\sim T_s^2\rho^2 eV/\hbar$. 
The dephasing rate is about the same.
The measurement rate given by 
$\hbar \Gamma_m\sim T_s^4\rho^2 B^2/(T_0^2 e V)$ is 
smaller than the spin-flip rate. 
Hence, tunneling measurement of spin is impossible in the normal state. 

Thermal quasiparticles cause spin flips as the electrons do in the 
normal state. The demolition rate is
proportional to the fraction of quasiparticles. At $T\ll\Delta_0$ 
it is exponentially small, $\hbar \Gamma_s^{\dagger}\lesssim
T_s^2\rho^2  eV\exp(-\Delta_0 /T)$. 

In this paper we have considered spin $S=1/2$.
We note that also for arbitrary $S$ the current operator
$\langle \hat{I}(t)\rangle_e$ is a linear combination of operators
$\hat{S}_z^n$, $0\leq n\leq 2S$. It depends on the spin initial state
and differs for the states $\langle \hat{S_z}\rangle=\pm S$
in contrast to the classical spin (discussed in Ref.~\onlinecite{Zh})
for which it depends only on $S_z^2$ because $\tilde{{\bf S}}(t)
\cdot\tilde{{\bf S}}(0)=(S^2-S_z^2)\cos\omega_Lt+S_z^2$.

We note, that other strategies of spin measurement could also be 
considered. E.g., a
Josephson junction with a spin could be inserted into a
superconducting ring, a voltage across the junction could be induced
by changing magnetic flux across the ring and the alternating flux
induced by Josephson oscillations could be measured by a secondary circuit 
coupled inductively to the superconducting ring. This will be studied 
elsewhere.

In conclusion, we have shown that the nondemolition measurements of
a quantum spin are possible using the Josephson oscillations. The average 
amplitude of Josephson oscillations depends on the initial state of the spin
after the tunneling or voltage is switched on and the voltage exceeds the 
dissipative threshold. This amplitude can be found measuring the dc-current 
in the circuit with a resistor. 
Use of the singlet-pair tunneling as an intermediate nondissipative system 
allows us to obtain information on the initial value of the $z$ component of 
the spin without flipping the spin. Thus a QND tunneling measurement of spin 
is performed.

The authors thank I. Martin, D. Averin, G. Ortiz, A. Balatsky and Yu. Galperin
for useful discussions. This work was supported by the
US DOE and by the ESF ``Vortex'' Program.

\end{multicols}

\end{document}